\documentclass[reprint,showpacs,amsmath,amssymb,aps]{revtex4-1}

\usepackage{graphicx}
\usepackage{dcolumn}
\usepackage{amsfonts}
\usepackage{amsmath}
\usepackage{amssymb}
\usepackage{bm}
\usepackage[utf8]{inputenc}
\usepackage[usenames]{xcolor}

\begin{document}

\preprint{APS/123-QED}
\title{A semi-analytical solver for the Grad-Shafranov equation}

\author{D. Ciro}
 \email{dciro@if.usp.br}
\author{I. L. Caldas}
 \email{ibere@if.usp.br}
\affiliation{Departamento de Física Aplicada, Universidade de São Paulo, 05508-090, São Paulo, Brazil.}

\begin{abstract}
In toroidally confined plasmas, the Grad-Shafranov equation, in general a non-linear PDE, describes the hydromagnetic equilibrium of the system.
This equation becomes linear when the kinetic pressure is proportional to the poloidal magnetic flux and the squared poloidal current is a quadratic function of it.
In this work, the eigenvalue of the associated homogeneous equation is related with the safety factor on the magnetic axis, the plasma beta and the Shafranov shift, then, the adjustable parameters of the particular solution are bounded through physical constrains.
The poloidal magnetic flux becomes a linear superposition of independent solutions and its parameters are adjusted with a non-linear fitting algorithm.
This method is used to find hydromagnetic equilibria with normal and reversed magnetic shear and defined values of the elongation, triangularity, aspect-ratio, and X-point(s).
The resultant toroidal and poloidal beta, the safety factor at the $95\%$ flux surface and the plasma current are in agreement with usual experimental values for high beta discharges and the model can be used locally to describe reversed magnetic shear equilibria.

\end{abstract}

\maketitle

\section{Introduction}
The solution of the Grad-Shafranov equation~\cite{schluter1957,grad1958,shafranov1958} provides the magnetic field, the current density, and the kinetic pressure inside an axisymmetric plasma in hydromagnetic equilibrium.
Having analytical solutions to this equation is convenient to configure physical equilibria as a basis for theoretical studies of transport, waves, and stability.
It also allows to estimate the external magnetic field configuration necessary to confine a toroidal plasma with specified parameters~\cite{shafranov1972}.

The Grad-Shafranov equation is an elliptic PDE for the poloidal magnetic flux $\psi$ that labels the magnetic surfaces in an axisymmetric plasma equilibrium. The equation contains two \emph{arbitrary} functions $p(\psi)$ and $F(\psi)$ that specify the dependence of the kinetic pressure and the poloidal plasma current on the magnetic flux $\psi$.

Accordingly, the Grad-Shafranov equation is, in general, a nonlinear PDE and its solution rely on numerical methods.
However, for some choices of the arbitrary functions the equation becomes linear and separable, and the boundary value problem can be solved by superposition of independent solutions.
Various classes of analytical solutions have been introduced along the years~\cite{solovev1968,maschke1973,zheng1996,ludwig1997,mccarthy1999,guazzoto2007,cerfon2010}, usually involving linear and quadratic dependences of $p$ and $F^2$ on $\psi$.
These choices impose some inherit restrictions on the possible current density profiles, and relations between the physical parameters, but, in general, provide good magnetic topology and safety factor profiles.

In this work we study a class of exact solutions resulting when the pressure is a linear function of $\psi$ and the squared poloidal current is a quadratic function of $\psi$~\cite{mccarthy1999}.
These solutions are characterized by an eigenvalue that is related with the equilibrium parameters of an axisymmetric plasma, specifically, the Shafranov shift $\Delta$, the safety factor at the magnetic axis $q_0$ and the fraction of diamagnetic reduction of the toroidal field.
This relation allow us to establish the limits of the model and its parameters, and, using physical arguments, we define a region of consistent solutions in the parameters space.

Then, we employ the analytical solutions of the Grad-Shafranov equation to build a predictive solver that allows to specify the geometrical properties of the toroidal plasma, namely, the aspect ratio $\epsilon$, triangularity $\delta$, elongation $\kappa$ and X-point, for a given set of physical parameters $\{\Delta,q_0,f,\beta\}$.

The treatment presented in this work potentiate the use of this class of solution for predictive equilibrium calculations and to our knowledge the relations introduced here have not been presented elsewhere and provide a valuable tool for the construction of analytical equilibria. The whole treatment was done in dimensionless variables, so that the results can be properly scaled to any case of interest using a couple of machine parameters, the major radius $R_0$ and the toroidal vacuum field $B_0$. 

The manuscript is organized as follows. In the Section~\ref{AnalSol} we present a short survey on hydromagnetic equilibrium with the model considered in this work and the analytical solutions to the Grad-Shafranov equation, then, in the Section~\ref{Parameters}, we study the relations between the physical parameters and the solution parameters. In the Section~\ref{Numerical}, we introduce a numerical method to solve the boundary value problem, and, in the Section~\ref{Results}, we give some examples of equilibrium calculations performed with this method and present our conclusions in the Section~\ref{Conclusions}.
\section{Analytical solutions}\label{AnalSol}

For ideal plasmas, the equilibrium between the kinetic and magnetic forces requires that
\begin{equation}\label{e01}
 \vec j\times\vec B = \nabla p,
\end{equation}
everywhere inside the plasma. Here, $p$, $\vec j$ and $\vec B$ are the kinetic plasma pressure, current density and magnetic field respectively. Assuming that the system is axisymmetric and using the Ampère's law the equilibrium problem is reduced to the Grad-Shafranov equation~\cite{schluter1957,grad1958,shafranov1958}
\begin{equation}\label{e02}
 R^2\nabla\cdot\left(\frac{\nabla\psi}{R^2}\right)=-\mu_0R^2\frac{dp}{d\psi}-F\frac{dF}{d\psi}=-\mu_0 R j_\phi.
\end{equation}
Here, $R$ is the distance to the symmetry axis and $\psi(R,z)$ is the poloidal magnetic flux, calculated though a disk of radius $R$ at the height $z$. The arbitrary function $p(\psi)$ represent the kinetic pressure at the level surface $\psi(R,z)=const.$, and $F(\psi)$ is the poloidal plasma current enclosed by that surface. The magnetic field lines lie on the magnetic surfaces $\psi(R,z)=const.$, that are also isobarics. Finally, the second equality in (\ref{e02}) relates the toroidal (azimuthal) current density with the arbitrary functions $p(\psi)$ and $F(\psi)$. 

From this point all the calculations are performed in dimensionless variables, so that the results can be scaled to any machine size and physical parameters. In dimensionless form, the Grad-Shafranov equation becomes
\begin{equation}\label{e03}
 \frac{\partial^2\bar\psi}{\partial x^2}-\frac{1}{x}\frac{\partial\bar\psi}{\partial x}+\frac{\partial^2\bar\psi}{\partial y^2}=-\varphi^2\left(\frac{\tilde{\beta}}{2}x^2\frac{d\bar p}{d\bar\psi}+\bar F\frac{d\bar F}{d\bar\psi}\right),
\end{equation}
where $\bar\psi(x,y)=\psi(R,z)/\psi_0$ is the normalized poloidal magnetic flux, and $\psi_0$ is the flux at the magnetic axis of the plasma. The variables $(x,y)=(R/R_0,z/R_0)$ are the normalized cylindrical coordinates, with $R_0$ the major radius of the plasma measured to the center of the poloidal cross section. The normalized arbitrary functions are $\bar p(\bar\psi)=p(\psi)/p_0$ and $\bar F(\bar\psi)=F(\psi)/R_0B_0$, where $p_0$ is the kinetic pressure at the magnetic axis and $B_0$ is the vacuum toroidal field at $R_0$. A characteristic beta was defined as $\tilde{\beta}=2\mu_0p_0/B_0^2$ and the parameter $\varphi=R_0^2B_0/\psi_0$ characterizes the ratio of toroidal and poloidal magnetic fluxes.

Setting to \emph{zero} the poloidal flux at the plasma edge, $\bar\psi$ grows monotonically towards \emph{one} at the magnetic axis. If the kinetic pressure is required to have a first order dependence of $\bar\psi$, it has to be simply
\begin{equation}\label{e04}
 \bar p(\bar\psi)=\bar\psi.
\end{equation}
This guarantees the vanishing of the pressure at the plasma edge, and a maximum value at the magnetic axis.
To set the form of the poloidal current $F(\psi)$, notice that the toroidal magnetic field has the form
\begin{equation}\label{e05}
 \bar B_\phi=\frac{\bar F(\bar \psi)}{x}.
\end{equation}
This field must tend to its vacuum form $B_\phi^v=1/x$ at the plasma edge where the plasma density vanishes.
Then, the poloidal current must satisfy $\bar F(0)=1$.
Requiring a second order dependence of $\bar F^2$ on $\bar\psi$ leads the general form
\begin{equation}\label{e06}
 \bar F^2(\bar\psi)=a\bar\psi^2+2b\bar\psi+1
\end{equation}
where the parameters $a$ and $b$ must be related to the equilibrium parameters. 
Using these arbitrary functions the Grad-Shafranov (\ref{e03}) equation takes the linear form
\begin{equation}\label{e07}
 \frac{\partial^2\bar\psi}{\partial x^2}-\frac{1}{x}\frac{\partial\bar\psi}{\partial x}+\frac{\partial^2\bar\psi}{\partial y^2}=-\varphi^2\left(\frac{\tilde{\beta}}{2}x^2+a\bar\psi+b\right).
\end{equation}
The poloidal flux is a superposition of an homogeneous solution $\psi_h$ and a particular solution $\psi_p$.
Assuming that $\psi_p$ depends only on even powers $x$, the particular solution may take the form
\begin{equation}\label{e08}
 \psi_p=-\frac{\tilde{\beta}}{2a}x^2-\frac{b}{a}.
\end{equation}
Other choices lead to infinite series expansions that unnecessarily complicate the analysis or provide solutions that appear in the homogeneous solution.

To solve the homogeneous equation, define $s^2=a\varphi^2$ for $a>0$ and $s^2=-a\varphi^2$ for $a<0$, so that $s$ is always a real number. Then we write the homogeneous equation as an eigenvalue problem
\begin{equation}\label{e09}
 \frac{\partial^2\psi_h}{\partial x^2}-\frac{1}{x}\frac{\partial\psi_h}{\partial x}+\frac{\partial^2\psi_h}{\partial y^2} = \mp s^2\psi_h.
\end{equation}
This can be solved by separation of variables with separation constants $\alpha,\gamma$ related to the eigenvalue through $\pm\alpha^2\pm\gamma^2=\pm s^2$.

For $a>0$ the homogeneous solution is a linear superposition of the following functions
\begin{equation}\label{e10}
 \psi_h(x,y)=
 \begin{cases}
  x[I_1,K_1(\alpha x)][\sin,\cos(\gamma y)],  &\gamma^2=\alpha^2+s^2\\
  [1,x^2][\sin,\cos(s y)],                    &\alpha=0\\
  x[J_1,Y_1(\alpha x)][\sin,\cos(\gamma y)],  &\gamma^2=s^2-\alpha^2\\
  x[J_1,Y_1(s x)][1,y],                       &\gamma=0\\
  x[J_1,Y_1(\alpha x)][\sinh,\cosh(\gamma y)],  &\gamma^2=\alpha^2-s^2
 \end{cases}
\end{equation}
Where $J_1,Y_1,K_1,I_1$ are the Bessel and Bessel modified functions of first order and we have used the abbreviated notation $[f_1,f_2(x)][g_1,g_2(y)]=c_1f_1(x)g_1(y)+c_2f_1(x)g_2(y)+c_3f_2(x)g_1(y)+c_4f_2(x)g_2(y)$, with $c_i$ arbitrary constants.
For the cases with $\alpha=0$ or $\gamma=0$ we use the dominant terms of the Bessel and harmonic functions for small arguments. This gives the same solutions that solving again the PDE (\ref{e09}) with a single separation constant.

Another possible solution can be obtained without separating variables and assuming spherical symmetry $\psi_h(x,y)=\psi_h(r)$ with $r=\sqrt{x^2+y^2}$. In this case the eigenvalue problem (\ref{e09}) becomes
\begin{equation}\label{e11}
 \frac{d^2}{d r^2}\psi_h(r)=-s^2\psi_h(r),
\end{equation}
with solutions
\begin{equation}\label{e12}
 \psi_h=\sin(sr), \cos(s r).
\end{equation}
This solution is relevant for the modern small aspect-ratio tokamaks and spheromaks, where the conducting chamber is D-shaped and the magnetic surfaces near the plasma edge are deformed accordingly. 

\begin{figure}[h]
 \centering
 \includegraphics[]{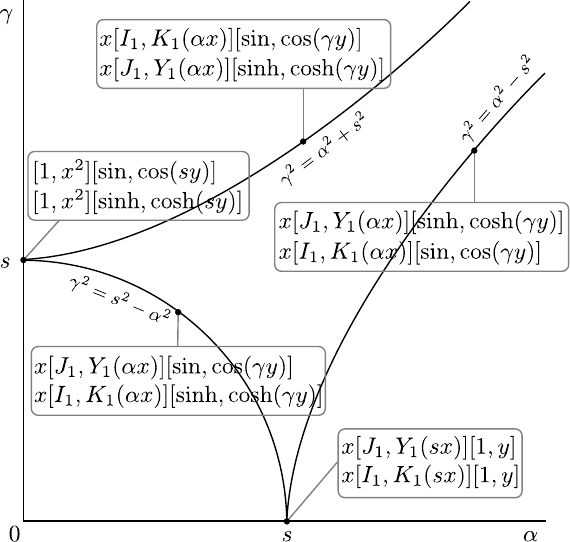}
 \caption{\label{fig1}Different forms of the solutions to the eigenvalue problem respect to the parameter space $\alpha-\gamma$. The upper functions on each box correspond to the solutions for $a>0$ and the bottom for $a<0$.}
\end{figure}

In analogy, for $a<0$ the homogeneous solution is a superposition of 
\begin{equation}\label{e13}
 \psi_h(x,y)=
 \begin{cases}
  x[J_1,Y_1(\alpha x)][\sinh,\cosh(\gamma y)],  &\gamma^2=\alpha^2+s^2\\
  [1,x^2][\sinh,\cosh(s y)], &\alpha=0\\
  x[I_1,K_1(\alpha x)][\sinh,\cosh(\gamma y)],  &\gamma^2=s^2-\alpha^2\\
  x[I_1,K_1(s x)][1,y], &\gamma=0\\
  x[I_1,K_1(\alpha x)][\sin,\cos(\gamma y)],  &\gamma^2=\alpha^2-s^2\\
 \end{cases}
\end{equation}
and the spherical solutions
\begin{equation}\label{e14}
 \psi_h=\sinh(s r),\cosh(s r).
\end{equation}
In Fig.~\ref{fig1} we arrange the solutions of (\ref{e09}) in the parameter space $\alpha-\gamma$, this illustrates the relation between the forms of the functions and the possible values of $\alpha,\gamma$.

In general, to solve a boundary value problem we will express the poloidal flux as
\begin{equation}\label{e15}
 \bar\psi(x,y)=-\frac{b}{a}-\frac{\tilde{\beta}}{2a}x^2+c_0G(s r)+\sum_{\alpha,\gamma}c_{\alpha,\gamma}B(\alpha x)H(\gamma y),
\end{equation}
where the form of the spherical solution $G$ and the functions $B$ and $H$ depend on the sign of $a$ (Fig.~\ref{fig1}). The values of the parameters $\alpha$, $\gamma$ and $s$ must be adjusted to satisfy the boundary conditions and the number of elements in the sum is, in principle, arbitrary. The superposition can also be expressed as an integral, but from the numerical point of view we only work with discrete values of $\alpha$ and $\gamma$.
\section{Equilibrium parameters}\label{Parameters}
Before dealing with the numerical method to solve the boundary value problem we need to establish relations between physical parameters and the parameters of the analytical solution (\ref{e15}).

From the toroidal magnetic field (\ref{e05}), the requirement for a diamagnetic plasma is $\bar F(\bar\psi)\lesssim 1$ on the plasma domain $0\leq\bar\psi\leq 1$. To keep track of this condition we define the constant $f=\bar F(1)$ and use it instead $b$ in the poloidal flux expansion. Using (\ref{e06}) we obtain $2 b=f^2-1-a$, and the squared poloidal current becomes
\begin{equation}\label{e16}
 \bar F^2(\bar\psi)=a\bar\psi(\bar\psi-1)-(1-f^2)\bar\psi+1.
\end{equation}
This form is more convenient to define explicitly the diamagnetic reduction of the toroidal magnetic field inside the plasma or its increase in paramagnetic cases. Using (\ref{e16}), the toroidal current density in units of $B_0/\mu_0R_0$ becomes
\begin{equation}\label{e17}
 \bar j_\phi=\frac{\varphi}{2 x}[\tilde{\beta} x^2+a(2\bar\psi-1)+f^2-1].
\end{equation}
Close to the magnetic axis, the safety factor can be approximated by
\begin{equation}\label{e18}
 q(\rho)=\frac{\rho \bar B_\phi}{(1+\Delta) \bar B_p(\rho)}.
\end{equation}
Here, $\rho,\Delta$ are the minor radius of the toroidal magnetic surface and the Shafranov shift in units of $R_0$, and  $\bar B_p,\bar B_\phi$ are the poloidal and toroidal components of the magnetic field in units of $B_0$. As $\rho\rightarrow 0$, (\ref{e18}) gives the exact value of the safety factor at the magnetic axis $q_0$. For $\rho$ small the poloidal magnetic field can be approximated by
\begin{equation}\label{e19}
 \bar B_p(\rho)=\frac{\bar j_0}{2}\rho,
\end{equation}
where $\bar j_0$ is the dimensionless toroidal current density at the magnetic axis. Setting $\bar\psi=1$ and $x=1+\Delta$ in (\ref{e17}) to replace $\bar j_0$ in (\ref{e19}), and replacing (\ref{e16}) in (\ref{e05}), the safety factor (\ref{e18}) becomes constant and equal to $q_0$
\begin{equation}\label{e20}
 q_0=\frac{4 f}{\varphi(1+\Delta)[\tilde{\beta}(1+\Delta)^2+a+f^2-1]}.
\end{equation}
Now, this relation is used to write $\varphi$ in terms of the other parameters, and is replaced in the eigenvalue equation 
\begin{equation}\label{e21}
s= \sqrt{|a|}\varphi.
\end{equation}
This gives $s$ in terms of $q_0$, $\Delta$, $f$ and the adjustable parameter $a$
\begin{equation}\label{e22}
 s = \frac{4 f\sqrt{|a|}}{q_0(1+\Delta)[\tilde{\beta}(1+\Delta)^2+a+f^2-1]}.
\end{equation}
For a predictive calculation we can set the values of the parameters $f,q_0,\Delta$ and $\tilde{\beta}$.
For instance, in a usual diamagnetic configuration $f\lesssim 1$, $q_0\gtrsim1$ and $\Delta\approx10^{-1}$.
To estimate the characteristic beta $\tilde{\beta}$, we use (\ref{e04}), and the definition of the toroidal beta $\beta_t=2\mu_0\langle p\rangle/B_0^2$, leading to
\begin{equation}\label{e23}
 \beta_t = \tilde{\beta}\langle\bar\psi\rangle,
\end{equation}
where $\langle\rangle$ denotes a volume average in the plasma domain. In a high-beta plasma the toroidal beta dominates the value of the total beta, $\beta\approx\beta_t$~\cite{freidberg}. Also, since $\bar\psi=0$ at the plasma edge and $\bar\psi=1$ on the magnetic axis, we can expect $\langle\bar\psi\rangle\approx0.5$. Then, for a given value of beta, we can approximate the characteristic beta by
\begin{equation}\label{e24}
 \tilde{\beta}\approx 2\beta
\end{equation}

\begin{figure}[h]
 \centering
 \includegraphics[]{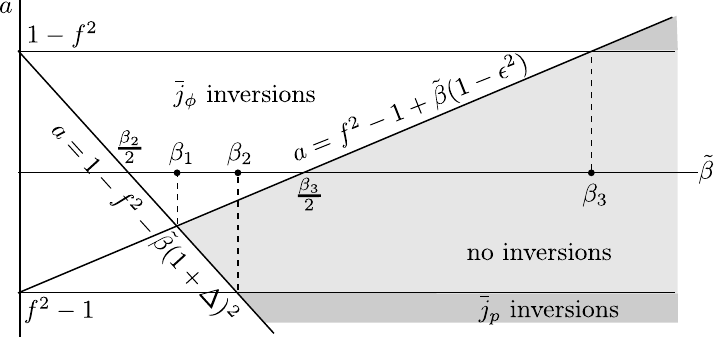}
 \caption{\label{fig2}The light region correspond to pairs $(a,\tilde{\beta})$ for which there are no toroidal or poloidal current density inversions. In the dark region the poloidal current density inverts but the toroidal do not.}
\end{figure}

To set the eigenvalue of the problem we need to know $a$ in (\ref{e22}). We can define $a$ as an adjustable parameter, but its range of allowed values must be established in a physical basis. To do this, we require the toroidal current density not to change its sign because it is mainly created by an inductive electric field. The signs of $j_\phi$ and $\psi_0$ must be the same, consequently the signs of $\bar j_\phi$ and $\varphi$ in (\ref{e17}) are the same, leading to the condition 
\begin{equation}\label{e25}
 \tilde{\beta} x^2+a(2\bar\psi-1)+f^2-1\geq0,
\end{equation}
in the whole plasma domain. This leads to
\begin{equation}\label{e26}
 1-f^2-\tilde{\beta}(1+\Delta)^2\leq a \leq \tilde{\beta}(1-\epsilon)^2+f^2-1,
\end{equation}
where $\epsilon$ is the aspect ratio of the plasma. Now, the condition for no poloidal current density inversions, comes from requiring the poloidal current $\bar F(\bar\psi)$ to be a monotonic decreasing function of $\bar\psi$, i.e. $\bar F'<0$. Using (\ref{e16}) we obtain the condition
\begin{equation}\label{e27}
 f^2-1<a<1-f^2.
\end{equation}
This conditions is useful to identify the parameters for poloidal current density inversions that may be required to describe the reversed magnetic shear equilibria emerging in situations with large bootstrap fractions.
Following the conditions (\ref{e27}) and (\ref{e26}) we can identify the regions of interest in the parameter space $a-\tilde{\beta}$ (Fig.~\ref{fig2}).

In Fig.~\ref{fig2} we depict the regions for poloidal current inversion $\bar j_p$, and no-inversions using the definitions
\begin{equation}\label{e28}
 \beta_2=2\frac{1-f^2}{(1+\Delta)^2} \mbox{ , }\beta_3=2\frac{1-f^2}{(1-\epsilon)^2}
\end{equation}
and $1/\beta_1=1/\beta_2+1/\beta_3$.
The restrictions over the allowed values of $a$ and $\tilde{\beta}$ defines through (\ref{e22}) the set of allowed eigenvalues $s$ that gives physical solutions to the boundary value problem. 
Given the form of the solutions (\ref{e15}) it is more convenient to adjust the eigenvalue $s$ than the parameter $a$. For this, we invert (\ref{e22}) to write $a$ in terms of $s$ and the physical parameters $q_0,\Delta,f,\tilde{\beta}$.
\begin{eqnarray}
 && a_\pm(s) = 1-f^2-\tilde{\beta}(1+\Delta)^2\pm\frac{8f^2}{q_0^2s^2(1+\Delta)^2}\times\nonumber\\
 && \left(1-\sqrt{1\pm\frac{q_0^2s^2(1+\Delta)^2}{4f^2}[1-f^2-\tilde{\beta}(1+\Delta)^2]}\right),\label{e29}
\end{eqnarray}
where $a_+(s)$ is valid for $a>0$ and $a_-(s)$ for $a<0$.

\begin{figure}[h]
 \centering
 \includegraphics[]{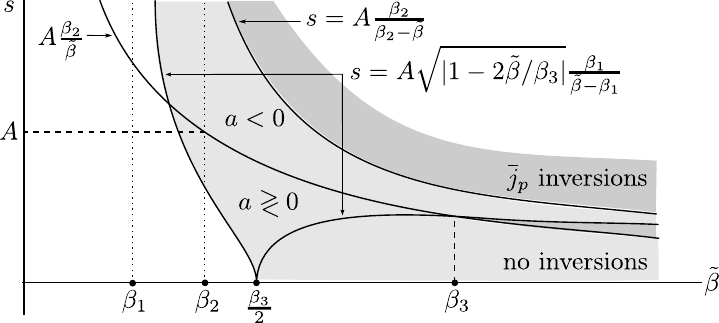}
 \caption{\label{fig3}Like in Fig.~\ref{fig2}, the light region correspond to pairs $(s,\tilde{\beta})$ for which there are no current density inversions and the darker region lead to poloidal current density inversions. As negative values of $s$ are not allowed the regions for $a<0$ and $a>0$ share a portion of the parameter space.}
\end{figure}

In Fig.~\ref{fig3}, we depict the regions of interest in the parameter space $s-\tilde{\beta}$ using the definition
\begin{equation}\label{e30}
 A=\frac{2}{q(1+\Delta)}\frac{f}{\sqrt{1-f^2}}.
\end{equation}
As in Fig.~\ref{fig2}, the region with $a<0$ is larger that the corresponding to $a>0$. Also, large values of $s$ lead to poloidal current inversions, except for a narrow region of $\tilde{\beta}$ between $\beta_1$ and $\beta_2$.
\section{Numerical method}\label{Numerical}
Now that we have characterized the equilibrium solutions respect to their position in the parameters space, we can develop a systematic method to build analytical solutions with some desired equilibrium properties.

Using $2b=f^2-1-a$ and (\ref{e29}) the poloidal magnetic flux (\ref{e15}) can be casted like
\begin{equation}\label{e31}
 \bar\psi(x,y)=\frac{1}{2}+\frac{1-f^2-\tilde{\beta} x^2}{2a_\pm(s)} + c_0G(sr)+\sum_{i}c_iB(\alpha_i x)H(\gamma_i y),
\end{equation}
with $\gamma_i=\gamma(\alpha_i,s)$ (see Fig.~\ref{fig1}).  The sign of $a$ determines its form in (\ref{e31}) and the functions $G, B$ and $H$ as explained in the Section~\ref{AnalSol}. The sign of $a$ is then kept unchanged during any optimization procedure that modifies the eigenvalue $s$, the coefficients $c_i$ and the parameters $\alpha_i$.

The Levenberg–Marquardt algorithm~\cite{marquardt1963} is used to adjust the linear and nonlinear parameters involved in this problem. This method gives good convergence for reasonable choices of the starting parameters. In general, the iterative process consists in the minimization of the error functional
\begin{equation}\label{e32}
 \varepsilon(\vec k)=\sum_{i=1}^N[\psi_i-\bar\psi(p_i,\vec k)]^2,
\end{equation}
where $p_i=(x_i,y_i)$ are points where we know the numerical values of the poloidal flux $\psi_i$, and $\bar\psi(p_i,\vec k)$ is our approximation to that value through (\ref{e31}) for a given set of $M$ parameters $\vec k=\{s,\{c_i\},\{\alpha_i\}\}$. The minimization of (\ref{e32}) is done by successive variations of $\vec k$,
\begin{equation}\label{e33}
 \vec k_1=\vec k_0+\vec\delta,
\end{equation}
where $\vec\delta$ must satisfy $\varepsilon(\vec k+\vec\delta)<\varepsilon(\vec k)$ and is obtained by solving the linear problem
\begin{equation}\label{e34}
 (J^TJ-\lambda I)\vec\delta(\lambda) = J^T[\vec y-\vec f(\vec k)].
\end{equation}
Here, $I$ is the $M\times M$ identity matrix, $\lambda$ is an adjustable parameter and the vectors are defined by
\begin{eqnarray}
 \vec y &=& (\psi_1,\psi_2,...,\psi_N)^T,\label{e35}\\
 \vec f(\vec k) &=& (\bar\psi(p_1,\vec k),\bar\psi(p_2,\vec k),...,\bar\psi(p_N,\vec k))^T\label{e36}.
\end{eqnarray}
$J$ is an $M\times N$ matrix with entries
\begin{equation}\label{e37}
 J_{i,j}=\frac{\partial\bar\psi(p_j,\vec k)}{\partial k_i},
\end{equation}
that in this case can be calculated analytically.
To update $\vec k$ and $\lambda$ we calculate the errors for $\vec\delta(\lambda)$ and $\vec\delta(\lambda')$ where $\lambda'=r\lambda$ and $0<r<1$. Then $\vec k$ and $\lambda$ are updated with the variation that gives the largest error reduction. If neither reduces the error we do $\lambda\rightarrow\lambda/r$ and repeat the previous step. Following this procedure we guarantee a rapid convergence far from the minimum and more refined steps close to it.

\begin{figure}[h]
 \centering
 \includegraphics[width=0.25\textwidth]{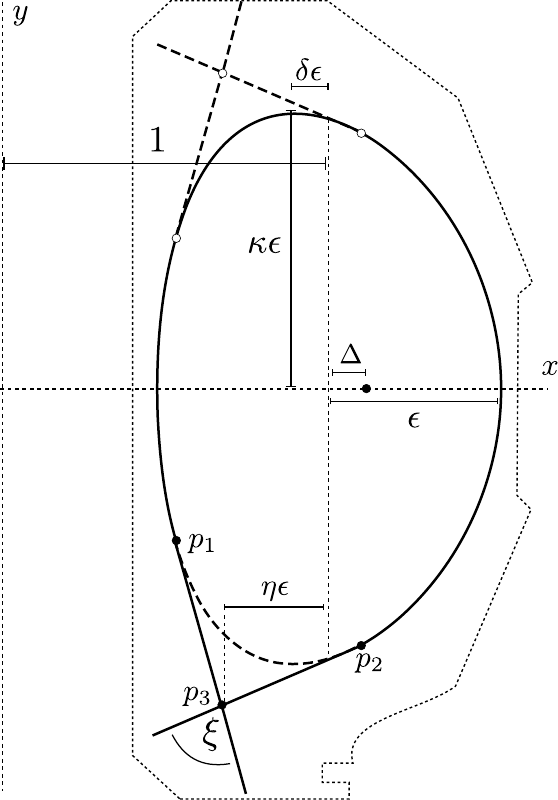}
 \caption{\label{fig4}The plasma edge is modeled by merging a D-shape with two tangent straight lines starting at $p_1$, $p_2$ and crossing at the X-point $p_3$. The center of the column is at $x=1$ and the magnetic axis is displaced by $\Delta$.}
\end{figure}

The points where the poloidal flux is known are on the plasma edge, where $\psi_i=0$ and the magnetic axis where $\psi_N=1$. To describe the plasma edge we can use a parametric equation containing the relevant geometry
\begin{eqnarray}
 x_b(\theta) &=& 1+\epsilon\cos(\theta+\alpha\sin\theta),\label{e38}\\
 y_b(\theta) &=& \epsilon\kappa\sin\theta,\label{e39}\\
 \delta &=& \sin\alpha\label{e40}.
\end{eqnarray}
This describes a D-shape with triangularity $\delta$, elongation $\kappa$ and minor radius $\epsilon$. In the case of a single or double null configuration we can trace straight lines that meet at the X-point at a distance $\eta\epsilon$ from the center with a desired angle $\xi$ (see Fig.~\ref{fig4}). For given values of $\xi$ and $\eta$, the positions of the X-point $p_3$, and the tangency points $p_1, p_2$ are uniquely determined and can be found by solving numerically an implicit equation.
\section{Results and discussion}\label{Results}
In the following, the numerical optimization described in the section~\ref{Numerical} will be used to find possible equilibrium configurations with realistic features in cases with normal and reversed magnetic shear.
\subsection{Normal shear equilibrium}
As a first example, the optimization algorithm is used to describe a shaped plasma with the parameters in Table~\ref{t1}.
\begin{table}[h]
\caption{\label{t1}Desired parameters} 
\begin{tabular}{cccccccc} 
 \toprule
  $\epsilon$ & $\kappa$ & $\delta$ & $\xi$ & $\eta$ & $\Delta$ & $q_0$ & $\beta$\\
  \hline
  $0.5$ & $1.5$ & $0.1$ & $0.4\pi$ & $0.6$ & $0.1$ & $1.1$ & $10\%$\\
  \hline
\end{tabular}
\end{table}
As we can not preset the value of $\beta$ in our method, we will set the value of $\tilde{\beta}$ following (\ref{e24}), assuming $\langle\bar\psi\rangle\approx 0.5$ for a usual discharge. From this we can estimate $\tilde{\beta}\approx 0.2$ as a starting guess for the method. We also start assuming a toroidal field reduction of $5\%$ relative to the vacuum value, i.e. $f\approx 0.95$. There is a close relation between the toroidal field fraction $f$ and the value of $\tilde{\beta}$, so we perform several runs for different combinations of $\tilde{\beta}$ and $f$ until we find the combination $(\tilde{\beta},f)=(0.26,0.97)$, leading to the best error reduction and magnetic topology for the chosen expansion of the flux $\bar\psi$.

To choose the basis we first set $a>0$ that corresponds to the solution (\ref{e10}), and leads to a maximum of the toroidal current density inside the plasma, (see (\ref{e17})), otherwise we could get a minimum which is only relevant in cases with reversed magnetic shear. The choice of the basis elements to expand the poloidal flux is somewhat intuitive. We start by choosing the functions on each branch of the Fig.~\ref{fig1}, then we turn on/off the different elements of the basis to see if the performance of the method is improved. After a few trials we keep the expansion that best minimizes the error, presenting the most physically relevant plasma profiles and topology of the magnetic surfaces. The resulting expansion is

\begin{eqnarray}
 && \bar\psi(x,y)=1/2+(1-f^2-\tilde{\beta} x^2)/2a(s)+(c_1+c_2 y)x Y_1(sx)\nonumber \\
 && +c_3\sin(s y)+c_4\cos(s y)+c_5\sin(sr)+c_6\cos(sr)\nonumber \\
 && +c_7xJ_1(\alpha_7 x)\sin(\gamma_7 y)+c_{8}xJ_1(\alpha_{8} x)\cos(\gamma_{8}y)\nonumber \\
 && +c_{9}xY_1(\alpha_{9} x)\sin(\gamma_{9} y)+c_{10}xY_1(\alpha_{10} x)\cos(\gamma_{10}y)\nonumber \\
 && +c_{11}xK_1(\alpha_{11} x)\sin(\gamma_{11} y)+c_{12}xK_1(\alpha_{12} x)\cos(\gamma_{12}y)\nonumber \\
 && +c_{13}xJ_1(\alpha_{13} x)\cos(\gamma_{13}y)+c_{14}xY_1(\alpha_{14} x)\cos(\gamma_{14}y).\label{e41}
\end{eqnarray}
with $r=\sqrt{x^2+y^2}$.
Using the restrictions (\ref{e26}),(\ref{e27}) over $a$, we were able to estimate the starting eigenvalue on $s\approx 3.0$, and the initial values of the starting parameters were chosen to be $\alpha_{7-10}=0.5 s$, $\alpha_{11,12}=3.5 s$ and $\alpha_{13,14}=0.3 s$.
These values evolve independently of $s$ during the optimization process, then, they will spread in the parameter space $\alpha-\gamma$. The initial values of the expansion coefficients $\{c_i\}$ are calculated by solving the linear problem of minimizing the error $\varepsilon(\{c_i\})$ for $s,\{\alpha_i\}$ fixed on the starting values.

\begin{figure}[h]
 \centering
 \includegraphics[]{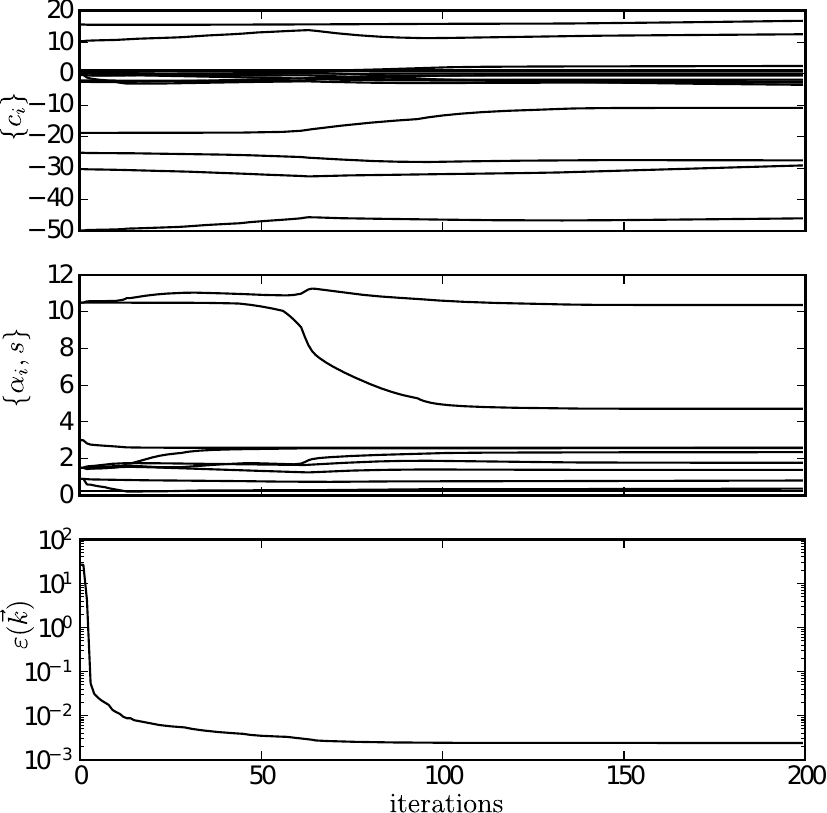}
 \caption{\label{fig5}Traces of the expansion coefficients $\{c_i\}$ and the nonlinear parameters $s,\{\alpha_i\}$ for $200$ iterations of the minimization method. The eigenvalue $s$ stabilizes at $2.69$ and the error at $2.4\times 10^{-3}.$}
\end{figure}

In Fig.~\ref{fig5} we can see the evolution of the solution parameters as the error is reduced from $26.7$ to $2.4\times 10^{-3}$ in 200 iterations of the method, when the parameters and the error do not change significantly the run ends.

\begin{figure}[h]
 \centering
 \includegraphics[]{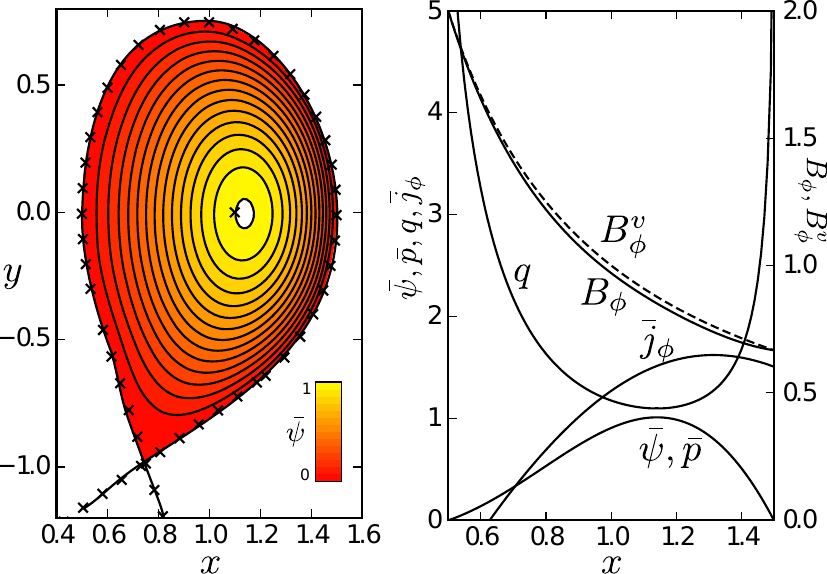}
 \caption{\label{fig6}In the left, the poloidal magnetic flux contours show good agreement with the imposed control points ($\times$), at the edge and magnetic axis. In the right, the resulting profiles of $q(x),B_\phi(x),\bar\psi(x),\bar p(x)$ for a transversal cut $y=0$, behave as expected for a divertor discharge, and the vacuum magnetic field $\bar B_\phi^{v}=1/x$ stands above $B_\phi$, illustrating the diamagnetic effect of the poloidal plasma current.}
\end{figure}

After the minimum is reached and we are satisfied with the plasma shape we can calculate the relevant plasma profiles.
In Fig.~\ref{fig6}-left we depict the resulting topology of the magnetic surfaces and the $50$ control points used in the method, $\psi_{1-49}=0$ for the plasma edge and $\psi_{50}=1$ at the magnetic axis.
The plasma edge is in good agreement with the desired shape and the magnetic surfaces behave as expected with the magnetic axis slightly displaced from the desired position.

We use (\ref{e05},\ref{e16},\ref{e29}) to calculate $\bar B_\phi$ and compare with the vacuum toroidal field $\bar B_\phi^{v}=1/x$. In Fig.~\ref{fig6}-right we can see the reduction of the toroidal field due to the diamagnetic effect controlled by $f$.
The pressure profile is by definition the same of $\bar\psi$ and the toroidal current density $\bar j_\phi$ is calculated with (\ref{e17}) using the relations (\ref{e21},\ref{e22},\ref{e29}).

For $\tilde{\beta}=0.26$ the obtained eigenvalue is slightly outside the shaded region in Fig.~\ref{fig3}, accordingly, there is a moderated inversion in $j_\phi$ coming from the inherit restrictions of the model, and, in addition, $\bar j_\phi$ does not vanish at the edge in the low field side.

To understand the $j_\phi$ profile, notice in (\ref{e17}), that the toroidal current density is not constant at the plasma edge, where $\bar\psi=0$, but changes with the radial distance $x$. Consequently, the current density can not vanish in the low and high field side simultaneously. This is a direct consequence of the choice of the profiles of $p(\psi)$ and $F(\psi)$ that makes linear the Grad-Shafranov equation~(\ref{e03}), and such unphysical behavior is acceptable when working with analytical models.

The safety factor $q(\bar\psi)$ is the constant ratio $d\phi/d\vartheta$ of the toroidal and poloidal angles subtended by the magnetic line as it wanders over its invariant surface. The poloidal angle $\vartheta$, is not uniform in the Cartesian space $\{x,y\}$, but the toroidal one is, so, we can calculate $q(\bar\psi)$ by following the magnetic line until it completes a full poloidal cycle, then we use
\begin{equation}\label{e42}
 q(\bar\psi)=\frac{\Delta\phi}{2\pi}.
\end{equation}
Doing this for a set of initial conditions in the line $y=0$ we get the $q$-profile in Fig.~\ref{fig6}-right. The value $q_{min}=1.096$ is in close agreement with the desired $q_0=1.1$, and the safety factor at the $95\%$ flux surface (in our case $\bar\psi=0.05$), is $q_{95}=4.515$. This is a typical value for a divertor discharge and was not preset in the analytical solution, but comes naturally from the elements of the expansion and the boundary conditions. The volume averaged poloidal flux is $\langle\psi\rangle=0.45$, then, from (\ref{e23}) the toroidal beta is $\beta_t=11.7\%$. We also calculate the poloidal beta given by $\beta_p=2\mu_0\langle p\rangle/\bar B_p^2$ where $\bar B_p=\mu_0I_0/2\pi a\kappa$, $a$ is the minor radius, $\kappa$ the elongation and $I_0$ the plasma current. In dimensionless variables we can write this like
\begin{equation}\label{e43}
 \beta_p=\tilde{\beta}(2\pi\epsilon\kappa)^2\frac{\langle \bar\psi\rangle}{\bar I_0^2},
\end{equation}
where $\bar I_0=\int\int\bar j_\phi(x,y)dxdy$ is the plasma current in units of $R_0B_0/\mu_0$, namely, the current used to create the vacuum toroidal field. Replacing our values we obtain $\beta_p=2.86$, then we can calculate the total beta using $\beta^{-1}=\beta_t^{-1}+\beta_p^{-1}$, and we obtain $\beta=11.2\%$ that is just $1.2\%$ above the desired value.

\begin{table}[h]
\caption{\label{t2}Obtained parameters} 
\begin{tabular}{ccccccccc}
 \toprule
  $\Delta$ & $\tilde{\beta}$ & $f$ & $q_{min}$ & $q_{95}$ & $\beta_t$ & $\beta_p$ & $\beta$ & $\bar I_0$ \\
  \hline
  $0.14$ & $0.26$ & $0.97$ & $1.096$ & $4.515$ & $11.7\%$ & $2.86$ & $11.2\%$ & $1.048$\\
  \hline
\end{tabular}
\end{table}

Table~\ref{t2} summarizes the results presented for this equilibrium. For the chosen poloidal flux expansion (\ref{e41}), the optimization method led to good agreement with the expected values of Table~\ref{t1}, and reasonable prediction for the values of the plasma current $\bar I_p$, the poloidal beta $\beta_p$, total beta $\beta$, and the $95\%$ flux surface safety factor.

\subsection{Reversed shear equilibrium}
In plasmas with high bootstrap fraction the current density profile is fundamentally changed, presenting a central minimum, and maximum off-axis. This behavior, leads to a non-monotonic safety factor profile with maximum at the magnetic axis and minimum off-axis. In divertor discharges the minimum in $q$ is reinforced by the growth of $q$ to the plasma edge, where it diverges.
For this case we choose a double null equilibrium with the parameters in the Table~\ref{t3}. 
\begin{table}[h]
\caption{\label{t3}Desired parameters} 
\begin{tabular}{cccccccc} 
 \toprule
  $\epsilon$ & $\kappa$ & $\delta$ & $\xi$ & $\eta$ & $\Delta$ & $q_0$ & $\beta$\\
  \hline
  $0.37$ & $1.7$ & $0.3$ & $0.4\pi$ & $0.77$ & $0.06$ & $4.0$ & $7\%$\\
  \hline
\end{tabular}
\end{table}
In analogy to the previous case, we start with the guess $\tilde\beta=0.14$ and $f=0.95$, and perform several runs changing these values for a given choice of the poloidal flux expansion.
The central reversed magnetic shear was obtained for $f\gtrsim 1$, making the plasma slightly paramagnetic.
Values below \emph{one} led to non-monotonic safety factor profiles with several critical points.
The parameters that best minimized the error functional were $(\tilde\beta,f)=(0.14,1.03)$.

\begin{figure}[h]
 \centering
 \includegraphics[]{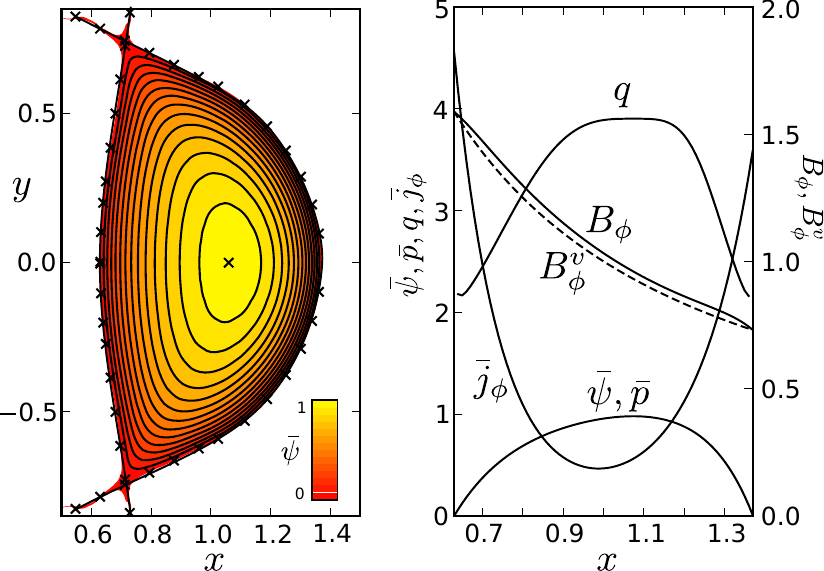}
 \caption{\label{fig7}In the left, the poloidal magnetic flux contours for a double null configuration and the control points ($\times$), at the edge and magnetic axis. In the right, the resulting profiles of $q(x),B_\phi(x),\bar\psi(x),\bar p(x)$ for a transversal cut $y=0$, in this case we have a hollow current profile and reversed magnetic shear. The vacuum magnetic field $\bar B_\phi^{v}=1/x$ stands below $B_\phi$, revealing a paramagnetic behavior of the plasma.}
\end{figure}

The hollow current profile requires $a<0$, corresponding to the solution (\ref{e13}) of the Grad-Shafranov equation.
Proceeding analogously to the previous case and using only even functions of $y$, we consider the following poloidal flux expansion
\begin{eqnarray}
 && \bar\psi(x,y)=1/2+(1-f^2-\tilde{\beta}x^2)/2a(s)+c_1\cosh(sr)\nonumber\\
 && +(c_2+c_3 x^2)\cosh(sy)+c_4xI_1(sx)+c_5xK_1(sx)\nonumber\\
 && +c_6I_1x(\alpha_6 x)\cos(\gamma_6 y)+c_7xK_1(\alpha_7 x)\cos(\gamma_7 y)\nonumber\\
 && +c_8xJ_1(\alpha_8 x)\cosh(\gamma_8 y)+c_9xY_1(\alpha_9 x)\cosh(\gamma_9 y)\nonumber\\
 && +c_{10}xI_1(\alpha_{10} x)\cosh(\gamma_{10} y)+c_{11}xK_1(\alpha_{11} x)\cosh(\gamma_{11} y)\nonumber\\
 && +c_{12}xJ_1(\alpha_{12} x)\cosh(\gamma_{12} y)+c_{13}xY_1(\alpha_{13} x)\cosh(\gamma_{13} y),\nonumber\\
 \label{e44}
\end{eqnarray}
the staring eigenvalue was established about $s\approx 8.0$, and the parameters were $\alpha_{6,7}=1.2s$, $\alpha_{8,9}=2.1 s$, $\alpha_{10,11}=0.6s$ and $\alpha_{12,13}=0.8$. In this case, the optimization method performed $300$ cycles and the error stabilized at $3.3\times 10^{-3}$. The eigenvalue stabilized close to the starting value at $s=8.245.$ The resulting profiles are presented in Fig.~\ref{fig7}.

For this equilibrium (Fig.~\ref{fig7}), the poloidal flux and kinetic pressure present a stronger drop at the plasma edge and the toroidal magnetic field is slightly increased ($3\%$) respect to its vacuum value, indicating a paramagnetic behavior.
The obtained current density presents a large hole at the plasma center and by inherit model restrictions it can not develop the off-axis maxima nor decrease to the plasma edge.

The safety factor profile presents the expected maximum at the magnetic axis with $q_{max}=3.9$, and develops an off-axis minimum near the plasma edge $q_{min}=2.16$, the minimum value, though consistent in value with equilibrium reconstructions of reversed magnetic shear discharges~\cite{diiid1995}, develops very close to the $95\%$ surface, causing an abrupt growth in $q$ close to the separatrix.
This is a consequence of the inability of the current density to decrease to the edge. Accordingly, this model is only able to represent a global reversed magnetic shear, and  should only be used locally to describe the central region of the plasma.

The magnetic axis is displaced by $1.3\%$ from the desired position and the toroidal and poloidal beta are $\beta_t=8.15\%, \beta_p=0.64$, leading to $\beta=7.22\%$ that is $0.22\%$ above the desired value.
The plasma current in units of $R_0B_0/\mu_0$ is $\bar I_0=1.414$.  This is a large value, e.g. if $R_0=1.6m$ and $B_0=2T$, $I_p=3.6MA$, and is caused by the unavoidable growth of the current density to the plasma edge.

\begin{table}[h]
\caption{\label{t4}Obtained parameters} 
\begin{tabular}{cccccccccc}
 \toprule
  $\Delta$ & $\tilde{\beta}$ & $f$ & $q_{max}$ & $q_{min}$ & $\beta_t$ & $\beta_p$ & $\beta$ & $\bar I_0$ \\
  \hline
  $0.074$ & $0.14$ & $1.03$ & $3.9$ & $2.16$ & $8.15\%$ & $0.64$ & $7.22\%$ & $1.414$\\
  \hline
\end{tabular}
\end{table}

For this equilibrium, the values of $q_{max}$, $q_{min}$, $\beta_t$, $\beta_p$ and $\beta$ are consistent with realistic situations, but the model is only able to reproduce the internal plasma behavior, and more flexible profiles for $q$ and $j_\phi$ requires higher powers of $\bar\psi$ on the source functions $\bar p(\bar\psi)$ and $\bar F(\bar\psi)$, and consequently, the solution of the full non-linear Grad-Shafranov equation.

\section{Conclusions}\label{Conclusions}
In this work we have identified the relevant parameters of a linear model of the Grad-Shafranov equation and related them with the plasma parameters of the hydromagnetic equilibrium, revealing the consistency regions in the parameter space.
For predictive calculations, the model parameters can be fixed and the poloidal magnetic flux becomes a linear superposition of the solutions to the linear Grad-Shafranov equation. This introduces a number of free parameters that can be adjusted numerically.

We have applied an optimization method to adjust the free parameters in a situation where the plasma edge and magnetic axis were established.
This method is able to produce single or double-null equilibrium configurations with realistic geometry and parameters. It achieves good magnetic topology, safety factor profiles, realistic values of $\beta$, and allows the explicit control the amount of diamagnetism or even paramagnetism in the plasma.
However, it has some inherit limitations in the current density profile due to the choice of the arbitrary functions.

The presented description of the plasma can be used globally to study the equilibrium in usual and high-$\beta$ discharges, and locally to describe the internal plasma in a reversed magnetic shear configuration. In both cases the convergence to the solution is good for reasonable choices of the basis functions in the poloidal flux expansion. 

The authors wish to thank Professors R.M.O. Galvão and Z.O. Guimarães-Filho for their useful discussions. This research was depeloped with the financial support of the Brazilian scientific agencies: CAPES, CNPq and the São Paulo Research Foundation (FAPESP), grants 2012/18073-1 and 2011/19269-11.

\bibliography{references}

\begin{thebibliography}{10}%
\makeatletter
\providecommand \@ifxundefined [1]{%
 \ifx #1\undefined \expandafter \@firstoftwo
 \else \expandafter \@secondoftwo
\fi
}%
\providecommand \@ifnum [1]{%
 \ifnum #1\expandafter \@firstoftwo
 \else \expandafter \@secondoftwo
\fi
}%
\providecommand \enquote [1]{``#1''}%
\providecommand \bibnamefont  [1]{#1}%
\providecommand \bibfnamefont [1]{#1}%
\providecommand \citenamefont [1]{#1}%
\providecommand\href[0]{\@sanitize\@href}%
\providecommand\@href[1]{\endgroup\@@startlink{#1}\endgroup\@@href}%
\providecommand\@@href[1]{#1\@@endlink}%
\providecommand \@sanitize [0]{\begingroup\catcode`\&12\catcode`\#12\relax}%
\@ifxundefined \pdfoutput {\@firstoftwo}{%
 \@ifnum{\z@=\pdfoutput}{\@firstoftwo}{\@secondoftwo}%
}{%
 \providecommand\@@startlink[1]{\leavevmode\special{html:<a href="#1">}}%
 \providecommand\@@endlink[0]{\special{html:</a>}}%
}{%
 \providecommand\@@startlink[1]{%
  \leavevmode
  \pdfstartlink
   attr{/Border[0 0 1 ]/H/I/C[0 1 1]}%
   user{/Subtype/Link/A<</Type/Action/S/URI/URI(#1)>>}%
  \relax
 }%
 \providecommand\@@endlink[0]{\pdfendlink}%
}%
\providecommand \url  [0]{\begingroup\@sanitize \@url }%
\providecommand \@url [1]{\endgroup\@href {#1}{\urlprefix}}%
\providecommand \urlprefix [0]{URL }%
\providecommand \Eprint[0]{\href }%
\@ifxundefined \urlstyle {%
  \providecommand \doi [1]{doi:\discretionary{}{}{}#1}%
}{%
  \providecommand \doi [0]{doi:\discretionary{}{}{}\begingroup
  \urlstyle{rm}\Url }%
}%
\providecommand \doibase [0]{http://dx.doi.org/}%
\providecommand \Doi[1]{\href{\doibase#1}}%
\providecommand \bibAnnote [3]{%
  \BibitemShut{#1}%
  \begin{quotation}\noindent
    \textsc{Key:}\ #2\\\textsc{Annotation:}\ #3%
  \end{quotation}%
}%
\providecommand \bibAnnoteFile [2]{%
  \IfFileExists{#2}{\bibAnnote {#1} {#2} {\input{#2}}}{}%
}%
\providecommand \typeout [0]{\immediate \write \m@ne }%
\providecommand \selectlanguage [0]{\@gobble}%
\providecommand \bibinfo [0]{\@secondoftwo}%
\providecommand \bibfield [0]{\@secondoftwo}%
\providecommand \translation [1]{[#1]}%
\providecommand \BibitemOpen[0]{}%
\providecommand \bibitemStop [0]{}%
\providecommand \bibitemNoStop [0]{.\EOS\space}%
\providecommand \EOS [0]{\spacefactor3000\relax}%
\providecommand \BibitemShut [1]{\csname bibitem#1\endcsname}%
\bibitem{schluter1957}%
  \BibitemOpen
  \bibfield{author}{%
  \bibinfo {author} {\bibfnamefont{V.}~\bibnamefont{Hain}}, \bibinfo {author}
  {\bibfnamefont{R.}~\bibnamefont{Lüst}},\ and\ \bibinfo {author}
  {\bibfnamefont{A.}~\bibnamefont{Schlüter}},\ }%
  \bibfield{journal}{%
  \bibinfo {journal} {Z. Naturforschg}\ }%
  \textbf{\bibinfo {volume} {12a}},\ \bibinfo {pages} {833} (\bibinfo {year}
  {1957})%
  \bibAnnoteFile{NoStop}{schluter1957}%
\bibitem{grad1958}%
  \BibitemOpen
  \bibfield{author}{%
  \bibinfo {author} {\bibnamefont{Grad}}\ and\ \bibinfo {author}
  {\bibnamefont{Rubin}},\ }%
  \bibfield{journal}{%
  \bibinfo {journal} {Proc. 2nd UN Conf. on the Peaceful Uses of Atomic
  Energy}\ }%
  \textbf{\bibinfo {volume} {31}},\ \bibinfo {pages} {190} (\bibinfo {year}
  {1958})%
  \bibAnnoteFile{NoStop}{grad1958}%
\bibitem{shafranov1958}%
  \BibitemOpen
  \bibfield{author}{%
  \bibinfo {author} {\bibfnamefont{V.}~\bibnamefont{Shafranov}},\ }%
  \bibfield{journal}{%
  \bibinfo {journal} {Sov. Phys. JETP}\ }%
  \textbf{\bibinfo {volume} {6}},\ \bibinfo {pages} {545} (\bibinfo {year}
  {1958})%
  \bibAnnoteFile{NoStop}{shafranov1958}%
\bibitem{shafranov1972}%
  \BibitemOpen
  \bibfield{author}{%
  \bibinfo {author} {\bibfnamefont{V.}~\bibnamefont{Shafranov}}\ and\ \bibinfo
  {author} {\bibfnamefont{L.}~\bibnamefont{Zakharov}},\ }%
  \bibfield{journal}{%
  \bibinfo {journal} {Nucl. Fusion}\ }%
  \textbf{\bibinfo {volume} {12}},\ \bibinfo {pages} {599} (\bibinfo {year}
  {1972})%
  \bibAnnoteFile{NoStop}{shafranov1972}%
\bibitem{solovev1968}%
  \BibitemOpen
  \bibfield{author}{%
  \bibinfo {author} {\bibfnamefont{L.}~\bibnamefont{Solov'ev}},\ }%
  \bibfield{journal}{%
  \bibinfo {journal} {Sov. Phys. JETP}\ }%
  \textbf{\bibinfo {volume} {26}},\ \bibinfo {pages} {400} (\bibinfo {year}
  {1968})%
  \bibAnnoteFile{NoStop}{solovev1968}%
\bibitem{maschke1973}%
  \BibitemOpen
  \bibfield{author}{%
  \bibinfo {author} {\bibfnamefont{E.~K.}\ \bibnamefont{Maschke}},\ }%
  \bibfield{journal}{%
  \bibinfo {journal} {Plasma Phys.}\ }%
  \textbf{\bibinfo {volume} {15}},\ \bibinfo {pages} {535} (\bibinfo {year}
  {1973})%
  \bibAnnoteFile{NoStop}{maschke1973}%
\bibitem{zheng1996}%
  \BibitemOpen
  \bibfield{author}{%
  \bibinfo {author} {\bibfnamefont{S.}~\bibnamefont{Zheng}}, \bibinfo {author}
  {\bibfnamefont{A.}~\bibnamefont{Wootton}},\ and\ \bibinfo {author}
  {\bibfnamefont{E.}~\bibnamefont{Solano}},\ }%
  \bibfield{journal}{%
  \bibinfo {journal} {Phys. Plasmas}\ }%
  \textbf{\bibinfo {volume} {3}},\ \bibinfo {pages} {1176} (\bibinfo {year}
  {1996})%
  \bibAnnoteFile{NoStop}{zheng1996}%
\bibitem{ludwig1997}%
  \BibitemOpen
  \bibfield{author}{%
  \bibinfo {author} {\bibfnamefont{G.}~\bibnamefont{Ludwig}},\ }%
  \bibfield{journal}{%
  \bibinfo {journal} {Plasma Phys. Control. Fusion}\ }%
  \textbf{\bibinfo {volume} {39}},\ \bibinfo {pages} {2021} (\bibinfo {year}
  {1997})%
  \bibAnnoteFile{NoStop}{ludwig1997}%
\bibitem{mccarthy1999}%
  \BibitemOpen
  \bibfield{author}{%
  \bibinfo {author} {\bibfnamefont{P.}~\bibnamefont{McCarthy}},\ }%
  \bibfield{journal}{%
  \bibinfo {journal} {Phys. Plasmas}\ }%
  \textbf{\bibinfo {volume} {6}},\ \bibinfo {pages} {3554} (\bibinfo {year}
  {1999})%
  \bibAnnoteFile{NoStop}{mccarthy1999}%
\bibitem{guazzoto2007}%
  \BibitemOpen
  \bibfield{author}{%
  \bibinfo {author} {\bibfnamefont{L.}~\bibnamefont{Guazzotto}}\ and\ \bibinfo
  {author} {\bibfnamefont{J.}~\bibnamefont{Freidberg}},\ }%
  \bibfield{journal}{%
  \bibinfo {journal} {Phys. Plasmas}\ }%
  \textbf{\bibinfo {volume} {14}},\ \bibinfo {pages} {112508} (\bibinfo {year}
  {2007})%
  \bibAnnoteFile{NoStop}{guazzoto2007}%
\bibitem{cerfon2010}%
  \BibitemOpen
  \bibfield{author}{%
  \bibinfo {author} {\bibfnamefont{A.~J.}\ \bibnamefont{Cerfon}}\ and\ \bibinfo
  {author} {\bibfnamefont{J.~P.}\ \bibnamefont{Freidberg}},\ }%
  \bibfield{journal}{%
  \bibinfo {journal} {Phys. Plasmas}\ }%
  \textbf{\bibinfo {volume} {17}},\ \bibinfo {pages} {032502} (\bibinfo {year}
  {2010})%
  \bibAnnoteFile{NoStop}{cerfon2010}%
\bibitem{freidberg}%
  \BibitemOpen
  \bibfield{author}{%
  \bibinfo {author} {\bibfnamefont{J.}~\bibnamefont{Freidberg}},\ }%
  \emph{\bibinfo {title} {Ideal Magnetohydrodynamics}}\ (\bibinfo {publisher}
  {Plenium Press, NY},\ \bibinfo {year} {1987})\ p.~\bibinfo {pages} {71}%
  \bibAnnoteFile{NoStop}{freidberg}%
\bibitem{marquardt1963}%
  \BibitemOpen
  \bibfield{author}{%
  \bibinfo {author} {\bibfnamefont{D.}~\bibnamefont{Marquardt}},\ }%
  \bibfield{journal}{%
  \bibinfo {journal} {J. Soc. Indust. Appl. Math.}\ }%
  \textbf{\bibinfo {volume} {11}},\ \bibinfo {pages} {431} (\bibinfo {year}
  {1963})%
  \bibAnnoteFile{NoStop}{marquardt1963}%
\bibitem{diiid1995}%
  \BibitemOpen
  \bibfield{author}{%
  \bibinfo {author} {\bibfnamefont{E.}~\bibnamefont{Strait}}, \bibinfo {author}
  {\bibfnamefont{L.}~\bibnamefont{Lao}}, \bibinfo {author}
  {\bibfnamefont{M.}~\bibnamefont{Mauel}}, \bibinfo {author}
  {\bibfnamefont{B.}~\bibnamefont{Rice}}, \bibinfo {author}
  {\bibfnamefont{T.}~\bibnamefont{Taylor}}, \bibinfo {author}
  {\bibfnamefont{K.}~\bibnamefont{Burrell}}, \bibinfo {author}
  {\bibfnamefont{M.}~\bibnamefont{Chu}}, \bibinfo {author}
  {\bibfnamefont{E.}~\bibnamefont{Lazarus}}, \bibinfo {author}
  {\bibfnamefont{T.}~\bibnamefont{Osborne}}, \bibinfo {author}
  {\bibfnamefont{S.}~\bibnamefont{Thompson}},\ and\ \bibinfo {author}
  {\bibfnamefont{A.}~\bibnamefont{Turnbull}},\ }%
  \bibfield{journal}{%
  \bibinfo {journal} {Phys. Rev. Lett.}\ }%
  \textbf{\bibinfo {volume} {75}},\ \bibinfo {pages} {4421} (\bibinfo {year}
  {1995})%
  \bibAnnoteFile{NoStop}{diiid1995}%
\end{thebibliography}%

\end{document}